\documentclass[11pt,a4paper]{article}

\usepackage[T1]{fontenc}
\usepackage[utf8]{inputenc}
\usepackage{lmodern}
\usepackage{microtype}
\usepackage[a4paper,margin=2.5cm]{geometry}
\usepackage{graphicx}
\usepackage{subcaption}
\usepackage{booktabs}
\usepackage{longtable}
\usepackage{multirow}
\usepackage{array}
\usepackage{amsmath,amssymb}
\usepackage{eurosym}
\usepackage{hyperref}
\usepackage{cite}

\hypersetup{colorlinks=true,linkcolor=blue,citecolor=blue,urlcolor=blue}

\title{Development and integration of the \texttt{NA64-DTC} automation controller for the CERN ``DESY Table'' motorized platform}

\author{%
Andrei Antonov$^{1,*}$\thanks{Corresponding author: \href{mailto:andrei.antonov@ge.infn.it}{andrei.antonov@ge.infn.it}} \and
Andrea Celentano$^{1,2}$ \and
Anna Marini$^{1}$ \and
Luca Marsicano$^{1}$\\[0.75em]
\normalsize $^{1}$Istituto Nazionale di Fisica Nucleare, Sezione di Genova \\
\normalsize $^{2}$Universit\`a degli Studi di Genova, Dipartimento di Fisica
}
\date{}

\begin{document}
\maketitle

\begin{abstract}
A remote automation controller \texttt{NA64-DTC} was developed, constructed, and integrated for the so-called ``DESY Table'' motorised platform, widely used in CERN East-Area and North-Area experimental installations. The device interfaces with the table manual control panel through a signal duplication connector, enabling remote operation without modifications to the original hardware. Button commands are emulated using opto-isolated switches, allowing seamless coexistence with the standard manual control system. The controller is based on an ESP32-C3 System-on-Module and hosts an \texttt{HTTP} server that provides a simple and flexible interface for experiment-specific integration and automation.

The system was successfully commissioned at CERN using the PS-T9 and SPS-H4 beamline facilities. A positioning accuracy of approximately $0.2$~mm and $0.3$~mm was achieved over the full movement range of the vertical and horizontal table axes respectively, with the residual positioning error dominated by the finite inertia of the platform. The device was operated continuously for more than two months without any observed failures or communication errors.

Although originally developed for the NA64 experiment, the proposed solution is applicable to any CERN installation employing the DESY Table platform. This work describes in detail the device technical design and operation, as well as the performances obtained during the commissioning.
\end{abstract}


\section{Introduction}
\label{sec:intro}

The CERN East-Area and North-Area beamlines can simultaneously deliver multiple particle beams with variable composition over a wide momentum range to several experimental installations hosting fixed-target experiments. These activities span from large-scale dedicated efforts  to smaller detector test and characterization measurements and proof-of-concept studies for novel experimental techniques (see e.g.~\cite{10.1063/1.5016162} for a comprehensive review). The key property of this experimental infrastructure is the large versatility in terms of the variety of the beams that can be transported to users installation, the instrumentation for control and diagnostic that can be installed and operated~\cite{Banerjee:2774716}, and finally the equipment and tools that can be employed by the different experimental installations. Among these, the so-called ``DESY Table'' motorized platform allows for precise and reproducible positioning of detectors and targets. Thanks to the high maximum load and the broad excursion range, the device is routinely used in fixed-target configurations for the characterization of the positional response of beam-absorbing detectors, mostly electromagnetic and hadronic calorimeters. During operations, frequent user intervention is required to change the platform configuration via a manual control panel, typically installed outside the experimental area, with the corresponding position solely available from local displays on this device. Among the various programs running at CERN North Area, the NA64 experiment routinely uses the DESY platform for precise positioning of the electromagnetic calorimeter (ECAL) employed as active target for its electron-beam, missing-energy program~\cite{NA64:2017vtt}. This happens both during the initial detector alignment and commissioning, as well as during the calorimeter energy calibration procedure~\cite{NA64:2025rib}. The limitations due to the manual operation of the current system, both in terms of the time required for platform movement and of the possible mistakes introduced by the user action,  motivated the development of a dedicated programmable, remote-control solution, the \texttt{NA64-DTC} (DESY Table Controller). This device could be integrated with the control system of any experiment running at CERN East-Area or North-Area, making it of general interest for the CERN experimental community. At CERN, a total of six DESY platforms are currently available, all sharing the same manual controller design.

This paper describes the design, implementation, and commissioning of the \texttt{NA64-DTC} device, interfacing with the DESY Table control panel via its built-in signal duplication connector. The controller emulates the electrical signals corresponding to panel button presses, achieving full
remote control without any hardware modifications to the original platform control system. Interaction with the device is achieved  through an embedded \texttt{HTTP} server: this general solution allows for a simple integration within any experiment-specific control system. So far, two complementary user interfaces have been developed, a generic browser-based Web UI for direct operator interaction, and an EPICS Input Output Controller (IOC)~\cite{Dalesio:1994qp,White:2019xlk} for integration with the NA64 slow-control subsystem, enabling scripted and interlocked operation.

The main technical contribution of this work lies in the overall system architecture developed to satisfy the unique constraints imposed by the DESY Table. In particular, the controller was designed to interface exclusively through the existing signal-duplication connector, requiring no modification of the original control electronics, while preserving the original manual operation and safety chain, allowing the automation controller to be installed or removed at any time without affecting the standard operation of the platform. The \texttt{NA64-DTC} also implements a closed-loop positioning system despite having no direct access to the motor drives, emulating the manual button commands to control the table while simultaneously reconstructing the platform position from the incremental encoder signals. These constraints required the development of a dedicated control architecture rather than the straightforward adoption of conventional industrial motion-control solutions typically used for motorized positioning platforms and other movable devices in accelerator environments, which typically assume direct access to the motor drives and are directly integrated into the system from the beginning of the design stage~\cite{nutter:icalepcs11-wepmn013,Makarov:2026bgg,Yue:2024sfw,Geng:2023ith}.

The remainder of this document is organised as follows.
Section~\ref{sec:design} describes the architectural design of the
controller, including an analysis of the control panel signal structure and
the rationale for the chosen approach. Section~\ref{sec:development} details
the hardware, firmware, and software development. Section~\ref{sec:commissioning}
reports on the device commissioning tests performed at CERN during the 2026 NA64 runs at the PS T9 and SPS H4 beamlines, while Section~\ref{sec:NA64} comments on the operational experience from the use of the \texttt{NA64-DTC} device during the NA64 2026 run at SPS. Conclusions are drawn in Section~\ref{sec:conclusions}.

\section{Materials and Methods}\label{sec:design}

\subsection{DESY Table characteristics and operation}
The DESY Table is a motorised platform designed for precision positioning in two axes (horizontal and vertical), with a maximum excursion of $\pm500$~mm in both directions, for a maximum load of about $10^{3}$~kg. DESY Table's motors are SEW-EURODRIVE R43W helical gear dual-speed motors which can work in two modes -- ``slow'' (31.5 RPM) and ``fast'' (126 RPM) -- and are equipped with Wachendorff WDGI 58B incremental encoders with 100 PPR (pulses per revolution) for position feedback. The incremental encoder is mounted on the motor assembly, while the integrated three-stages gearbox reduces the motor speed and increases the available torque before transmitting motion to the table. 

\begin{figure}[t]
    \centering
    \includegraphics[width=\linewidth]{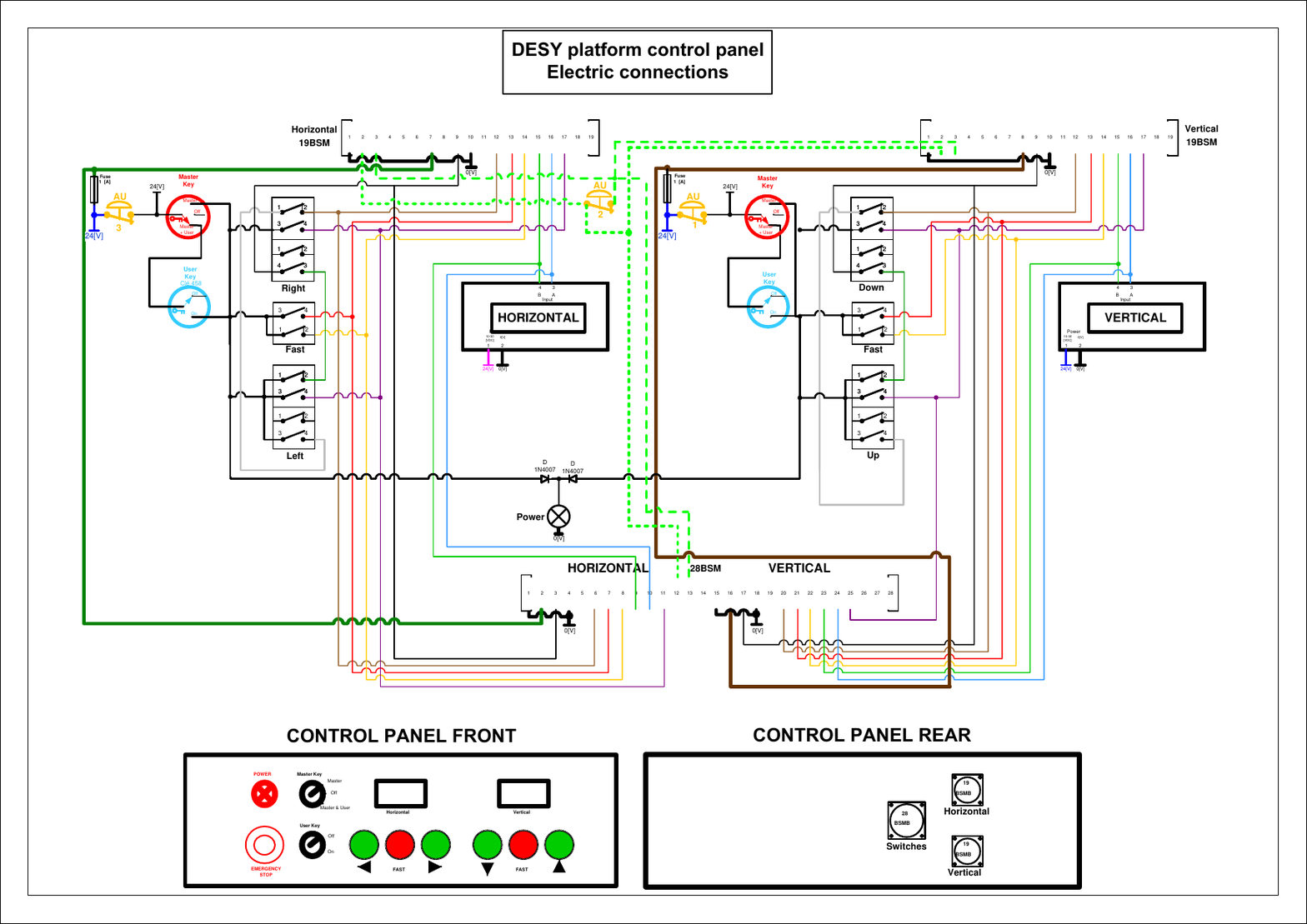}
    \caption{\label{fig:ctrl_panel} DESY Table control panel schematic (top part) and simplified schematic of front and rear buttons, keys, and displays (bottom part).}
\end{figure}

The platform is controlled via a dedicated control panel that allows for manual operation, including directional movement, speed selection, and position readout (see also Fig.~\ref{fig:ctrl_panel}). The control panel consists of a set of buttons for manual operation,
including directional controls (left, right, up, down), speed selection (fast) for both axes, and counters/displays for horizontal and vertical position readout. When operated in ``slow'' mode, the accuracy of the platform positioning is of about $0.5~$mm, mostly limited by the reproducibility of the operator's button press duration. The control panel also features an emergency stop button for safety and master and user keys for access, which can cut off the axes' 24~V DC power circuits (thick green and brown lines on schematic) from the control logic of the panel.
The rear of the control panel features two 1~A protection fuses (one per axis) and three Burndy connectors: two 19BSM for the axes' motor control and one 28BSM switches duplication connector,  which provides access to the electrical signals corresponding to the button presses (black, brown, orange, yellow and purple lines) and to the motors' incremental encoder outputs (blue and green lines) for the both axes. The remaining connections of the 28BSM connector are for the power supply and the emergency stop circuit. For reference, the full 28BSM pinout is presented in Table~\ref{tab:28bsm_pinout} in the Appendix~\ref{app:techDetails}. By interfacing with the 28BSM connector, it is possible to emulate button presses through opto-isolated switching, allowing for remote control without modifying the original hardware. 
This approach ensures that the standard DESY table setup is kept while enabling automation capabilities, so that the original control panel can still be used for manual operation if needed while the automation controller is connected, while the latter can be installed or removed without any physical alterations to the platform control panel.
The presence of the 24~V DC power supply lines on the 28BSM connector also allows the automation device to be powered directly from the DESY Table controller, eliminating the need for an external power source and simplifying the overall system design. 

\subsection{DESY Table automation controller development}\label{sec:development}

\subsubsection{Proposed device architecture}

\begin{figure}[t]
    \centering
    \includegraphics[width=0.9\linewidth]{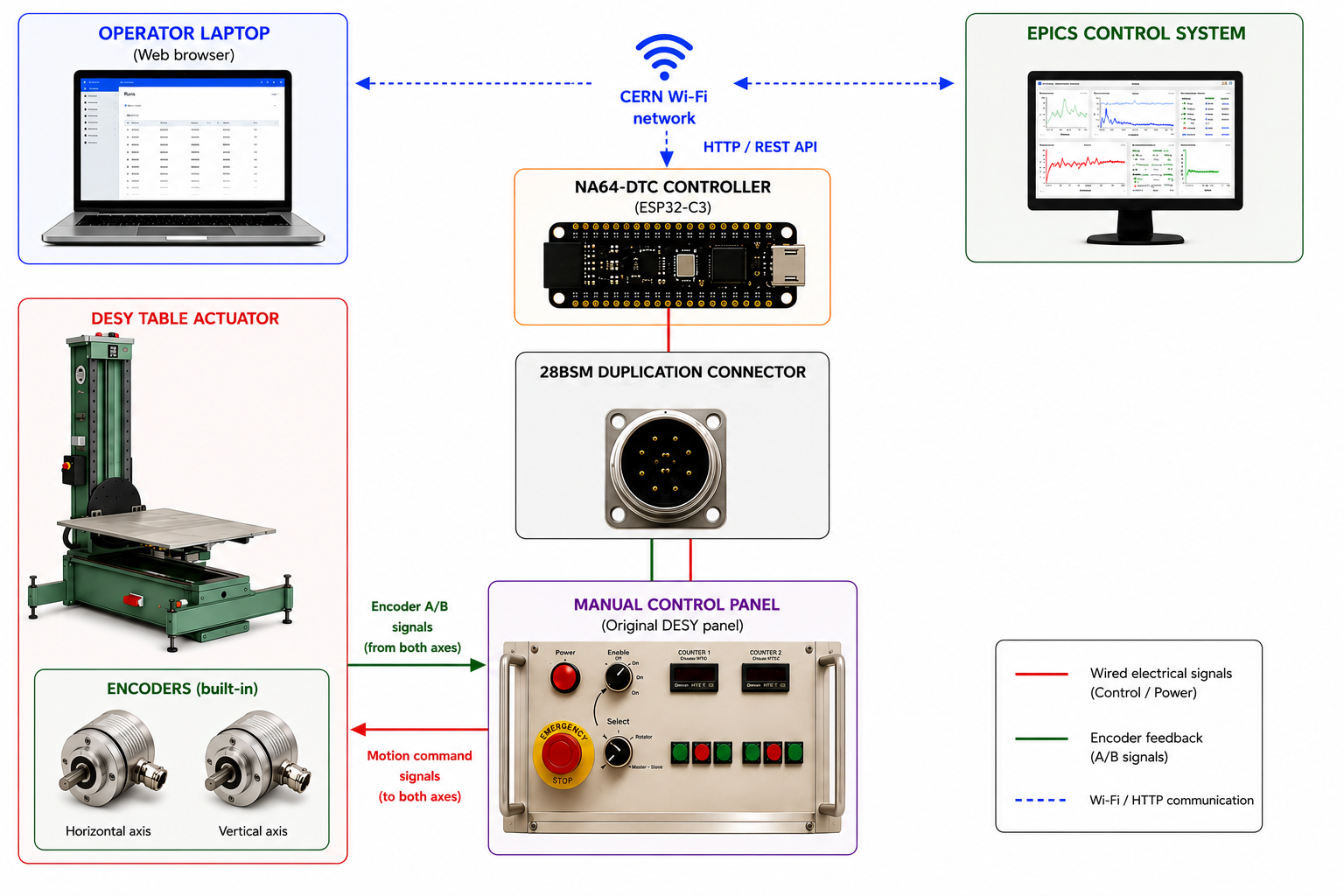}
    \caption{Overall architecture of the DESY-table remote control via \texttt{NA64-DTC} device. The DESY platform is electrically connected to the manual control panel. The \texttt{NA64-DCT} interfaces non-invasively with the latter through the 28BSM duplication connector.}
    \label{fig:NA64DST}
\end{figure}

The \texttt{NA64-DTC} was designed to fulfil the following requirements. To simplify connectivity and to allow for installation in any experimental setup, it should be a fully wireless device able to connect to the CERN Wi-Fi network
and to be controlled using standard network protocols. The power should be taken directly from the 24~V DC DESY platform line available from the control panel, with total consumption of less than 10~W to not overload it. It should be a compact device connected directly to the to 28BSM connector of the DESY table controller, that can be plugged or unplugged at any time without any interference with manual platform operations. Finally, since no limit switches, absolute-position sensors, or homing references are accessible through the external control interface, the \texttt{NA64-DTC} has to rely exclusively on the incremental encoder signals for position feedback, with travel limits implemented at software level. An overview of the overall system architecture is reported schematically in Fig.~\ref{fig:NA64DST}.

\begin{figure}
    \centering
    \includegraphics[width=0.8\linewidth]{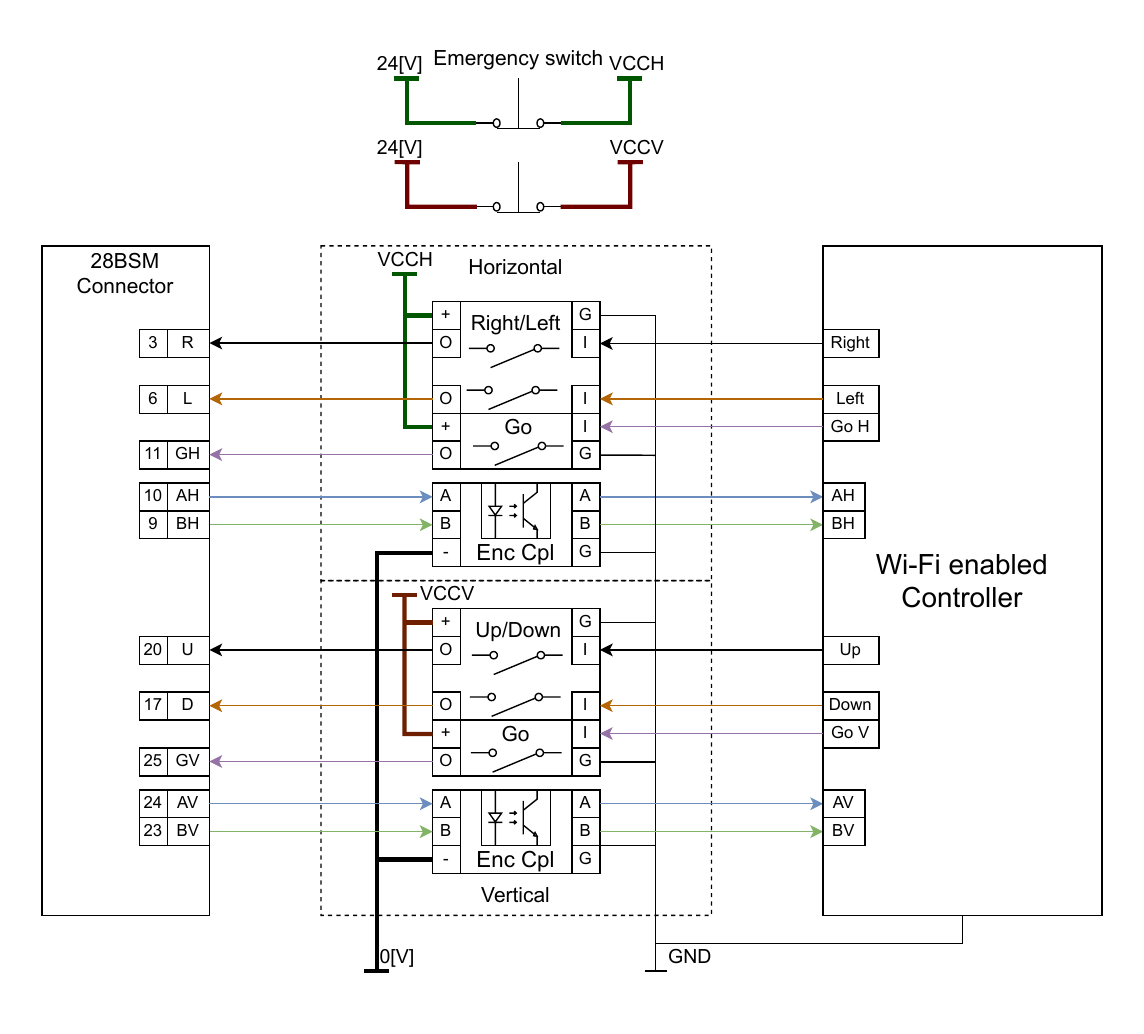}
    \caption{ \label{fig:desy_ctrl_struct} Simplified functional diagram of the \texttt{NA64-DTC} automation device.}
\end{figure}

The chosen architecture is based on a Wi-Fi enabled microcontroller which interfaces with the control panel via opto-isolated switches to emulate button presses and count the motor’s encoder pulses. The overall device structure is reported in Fig.~\ref{fig:desy_ctrl_struct}, showing from left to right the 28BSM connection pins, the opto-isolated switches on the \texttt{NA64-DTC} and the connections to the microcontroller. For each motor axis, three opto-isolated control switches are present, the first two (``Right/Left'' and ``Up/Down'') selecting the translation direction and the third (``Go'') enabling the movement. In the DESY control panel system (see Fig.~\ref{fig:ctrl_panel}), each pair of buttons selecting the motor direction is mutually exclusive, thanks to a feedback-loop connection, to ensure that no movement occurs if both are pressed simultaneously. This exclusion logic is implemented in the \texttt{NA64-DTC} controller firmware. Concerning the motor speed, the current design of the manual control panel is such that the corresponding button has two pairs of contacts, one being normally closed (``Slow'' mode), and the other normally open (``Fast'' mode). As a consequence, the ``Slow'' mode is always enabled by default by control panel - this forces the \texttt{NA64-DTC} design to not implement the speed-selection feature. The emergency switch shown on the diagram corresponds physically to the control panel emergency stop button. If this is pressed, it would disconnect power inputs from the opto-isolators to inhibit any remote-control of the platform. The emergency signal is also continuously monitored by the automation controller to implement its own emergency stop logic, as described in the next section. Finally, an optical-isolated phototransistor (``Enc Cpl'') is used to interface the controller with the signals from the two motors encoders to perform their incremental position readout. 

\subsubsection{Hardware design}

The simplicity of the device functionality suggests the use of a light-weight microcontroller with a bare-metal firmware
or a lightweight real-time operating system (RTOS). Ideally, the microcontroller should be equipped with a built-in
quadrature encoder interface to read the pulse counting from motor encoders. However, since the motors of the DESY platform
are spinning at low speed (52.5 counts/s in the ``Slow'' mode), this requirement can be lifted by implementing in the device a
software IRQ, thus allowing to choose from a broader range of devices. The more demanding part is the network connectivity for user interface, requiring the use of a  System-on-Chip (SoC) device with a built-in Wi-Fi interface, complemented by a TCP/IP stack implementation. 

The chosen solution was based on the recent ESP32-C3-WROOM-2 SoM, a low-power, versatile, low-cost device with embedded Wi-Fi connectivity~\cite{Espressif2025ESP32C3WROOM02}. This is based on a single-core RISC-V microcontroller with clock frequency up to 160 MHz, equipped with 400 KB of RAM and 4 MB of flash memory. The device also features a built-in USB-to-serial converter, which drastically simplifies the development and debugging process. The low power consumption of less than 2~W and the very small footprint makes this device ideal for the current application, allowing for a compact device that can be interfaced directly with the DESY platform manual controller. The ESP32-C3-WROOM-2 SoM does not include a built-in quadrature encoder
interface, requiring to implement the encoder pulse counting logic directly in the firmware through GPIO interrupts -- as previously commented, this does not present a limitation due to the low speed of the table motors.


\begin{figure}[t]
    \centering
    \begin{subfigure}[b]{0.9\linewidth}
        \centering
        \includegraphics[width=\linewidth]{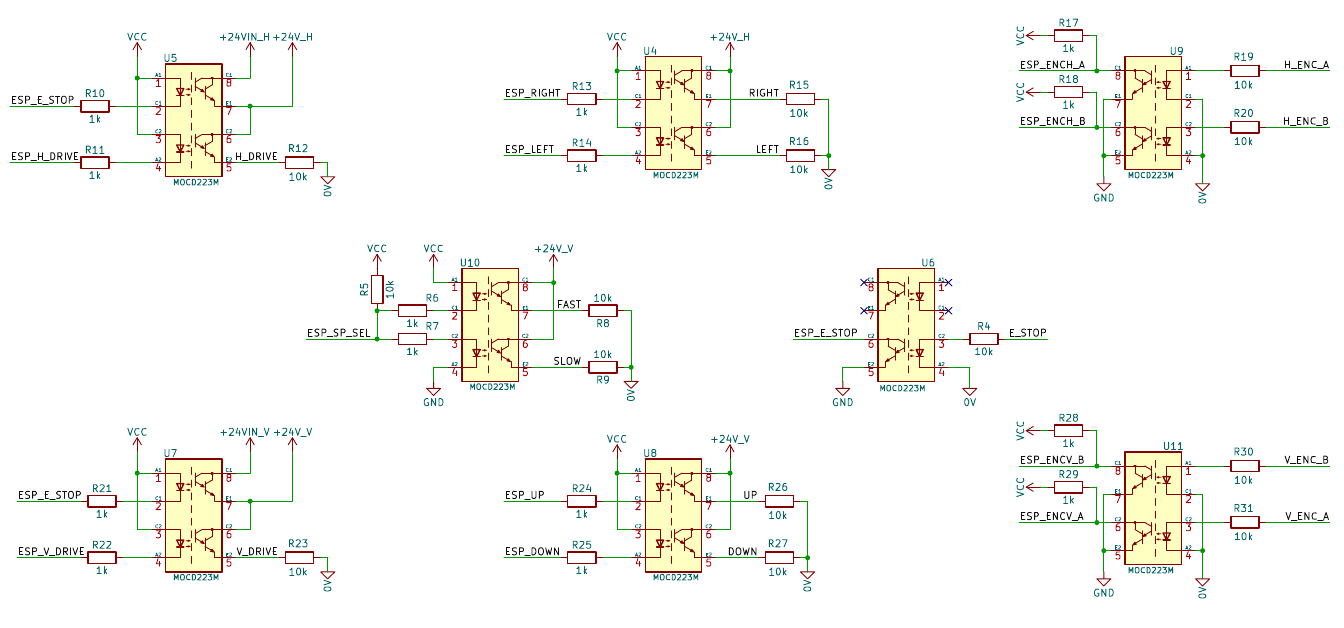}
        \caption{Opto-isolators: button emulation and encoder readout, E-Stop.}
        \label{fig:desy_ctrl_schematic_iso}
    \end{subfigure}

    \vspace{0.5em}

    \begin{subfigure}[b]{0.48\linewidth}
        \centering
        \includegraphics[width=\linewidth]{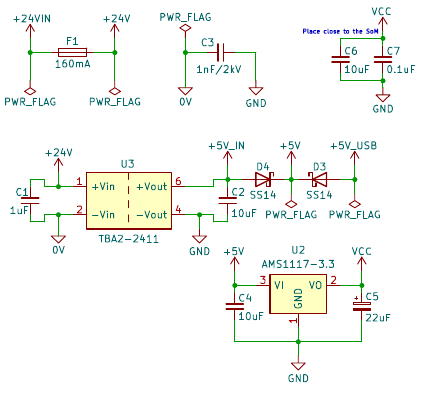}
        \caption{Power supply circuit.}
        \label{fig:desy_ctrl_schematic_power}
    \end{subfigure}
    \hfill
    \begin{subfigure}[b]{0.48\linewidth}
        \centering
        \includegraphics[width=\linewidth]{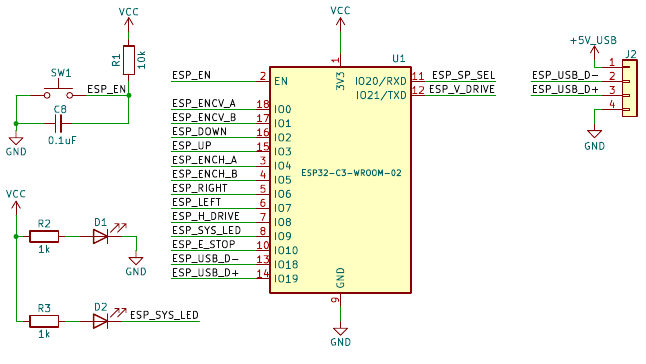}
        \caption{ESP32-C3-WROOM-2 SoM and its peripherals.}
        \label{fig:desy_ctrl_schematic_esp32}
    \end{subfigure}

    \caption{Schematic fragments of the \texttt{NA64-DTC} automation controller: (a) opto-isolators, (b) power supply, (c) ESP32-C3-WROOM-2 SoM and peripherals.}
    \label{fig:desy_ctrl_schematic}
\end{figure}

The device schematic is shown on Fig.~\ref{fig:desy_ctrl_schematic}. It consists of three main parts: the opto-isolator circuits for button emulation and encoder readout, the power supply circuit, and the ESP32-C3-WROOM-2 SoM (U1) with its peripherals.
To simplify the design, we decided to use one opto-isolator type for both button emulation and encoder readout. We opted for the MOCD223M dual-channel device, a low-cost and widely available Darlington output optocoupler with a very high current transfer ratio (CTR) of $500\%$ and a very low input current of about 1 mA. The relative long turn-off time typical of the Darlington output, of about 100 $\mu$s, is well within the specification of the current application, given the low frequency of the encoder pulses (52.5 Hz). 
\\
The device has in total 8 MOCD223M opto-isolators (5 for controls and 3 for readout). One (U6) is used to propagate the Emergency Stop signal from the +24~V power domain to the SoM power domain, allowing to implement the E-Stop control and monitoring logic inside the microcontroller\footnote{During normal operations, the E-Stop signal from the control panel is tied to +24~V through a NC contact of the corresponding button: when this is pressed, the signal is disconnected from the 24~V, the opto-isolator's channel switches off and the corresponding ESP32-C3's input goes pulled-up internally.}. To control the motors, six optocouplers in total are used:
\begin{itemize}
 \item the first one (U5/U7) is used for two different purposes: channel 1 uses E-Stop from the ESP32-C3 power domain to cut off power to all the other opto-isolators of the axis in case of an emergency stop condition, and channel 2 is used for the axis drive enable signal emulation;
\item the second one (U4/U8) is used for directional button emulation: channel 1 for Right/Up signals and channel 2 for Left/Down signals;
\item the third one (U9/U11) is used for encoder readout with channel 1 for encoder A output and channel 2 for encoder B output. It requires a low pull-up resistor (1 k$\Omega$) on the output side to ensure a reasonable switching time of the opto-isolator and to minimize the effect of the long turn-off time of the Darlington output.
\end{itemize}
Finally, even if the motor speed selection is currently unavailable in the \texttt{NA64-DTC} system due to the DESY manual controller limitations discussed previously, another optocoupler (U10) has been installed for this feature, to allow for a possible upgrade of the system in the future. All the button emulating outputs in +24~V power domain are pulled to ground through 10k$\Omega$ resistors to ensure that they are in a defined state when the opto-isolators are off. The control logic in the ESP32-C3 power domain is negative, which means that all the ESP32-C3 button emulating outputs are active low, they are connected to the cathodes of the opto-isolators' LEDs, and the anodes of the LEDs are connected to VCC. Negative logic is used to avoid unwanted switching during the power-up phase of the device, since the ESP32-C3's GPIOs might be pulled up during the boot process.

The power supply circuit consists of two primary subcircuits, a Traco Power's TBA2-2411 2~W isolated 24~V to 5~V DC/DC converter (U3) to supply the device from 28BSM connector during normal operation, and a direct, non-isolated 5~V connection supplied by USB-line during programming and debugging. To avoid conflicts between these two power rails, we implemented a simple power selection circuit based on two Schottky diodes, which allows the device to be powered from either source without any risk of back-feeding current into the other source. The 3.3~V power circuit for the ESP32-C3-WROOM-2 SoM is generated from the common 5~V line using an AMS1117-3.3 LDO (U2), which provides stable and low-noise power for the microcontroller and the RF-modem.

\begin{table}[t]
    \centering
      \caption{\texttt{NA64-DTC} IDC-20 connector pinout mapping to DESY platform manual controller 28BSM connector.}
    \label{tab:idc_pinout}
    \begin{tabular}{c|c|c|c|c|c}
        28BSM & Name & J1 & J1 & Name & 28BSM \\
        \hline
        20 & DOWN & 1 & 2 & 0V & 18 \\
        17 & UP & 3 & 4 & V\_ENC\_B & 24 \\
        25 & V\_DRIVE & 5 & 6 & V\_ENC\_A & 23 \\
        16 & +24V\_V & 7 & 8 & (V)FAST & 21 \\
        12 & E\_STOP & 9 & 10 & (H)FAST & 7 \\
        13 & +24V & 11 & 12 & (V)SLOW & 22 \\
        2 & +24V\_H & 13 & 14 & (H)SLOW & 6 \\
        11 & H\_DRIVE & 15 & 16 & H\_ENC\_A & 10 \\
        6 & LEFT & 17 & 18 & H\_ENC\_B & 9 \\
        3 & RIGHT & 19 & 20 & 0V & 4 \\
    \end{tabular}
  
\end{table}

\begin{figure}
    \centering
    \begin{subfigure}[b]{0.45\linewidth}
        \centering
        \includegraphics[width=\linewidth]{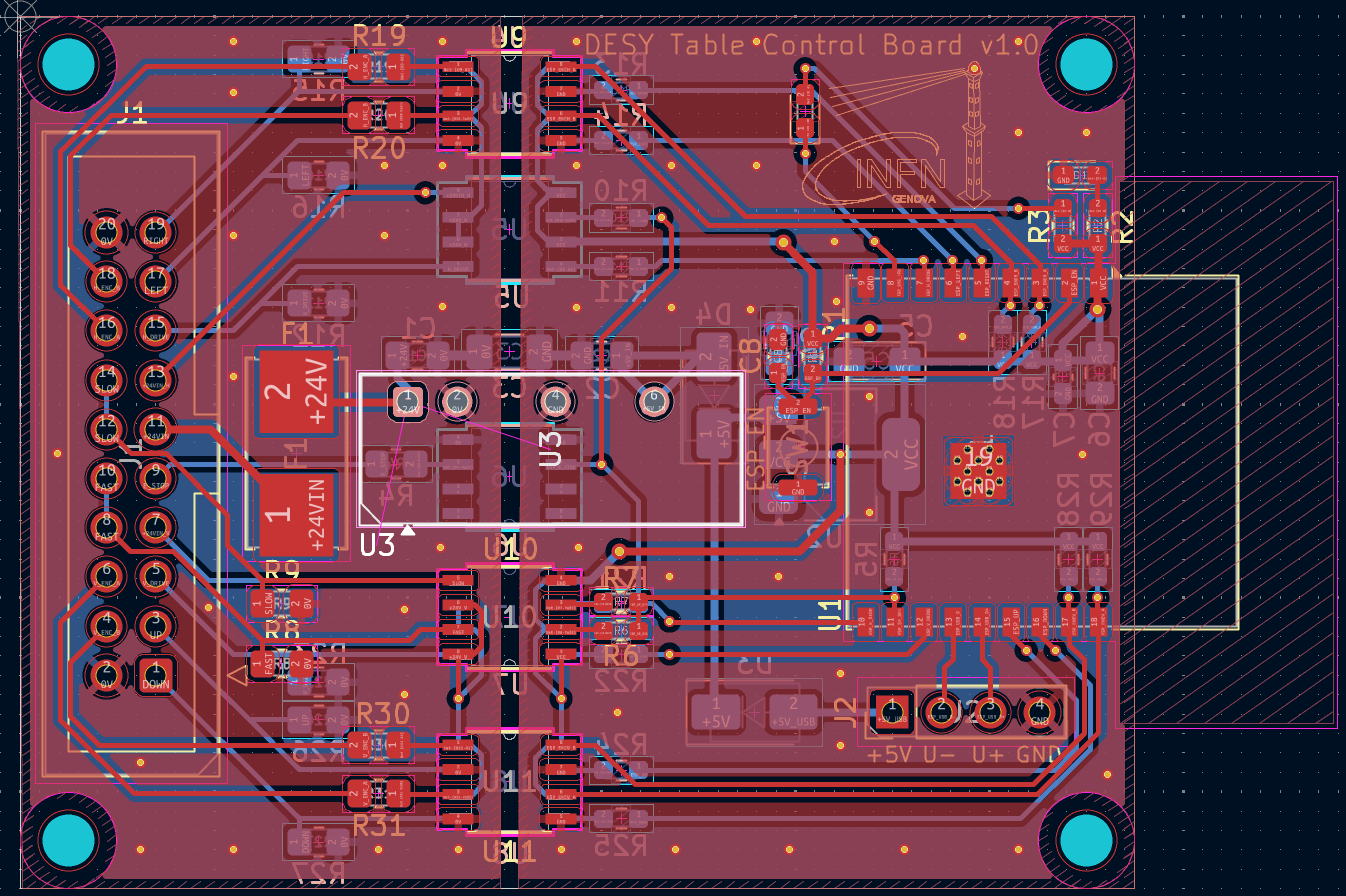}
        \caption{Traced PCB of the designed device.}
        \label{fig:desy_ctrl_pcb}
    \end{subfigure}
    \hfill
    \begin{subfigure}[b]{0.49\linewidth}
        \centering
        \includegraphics[width=\linewidth]{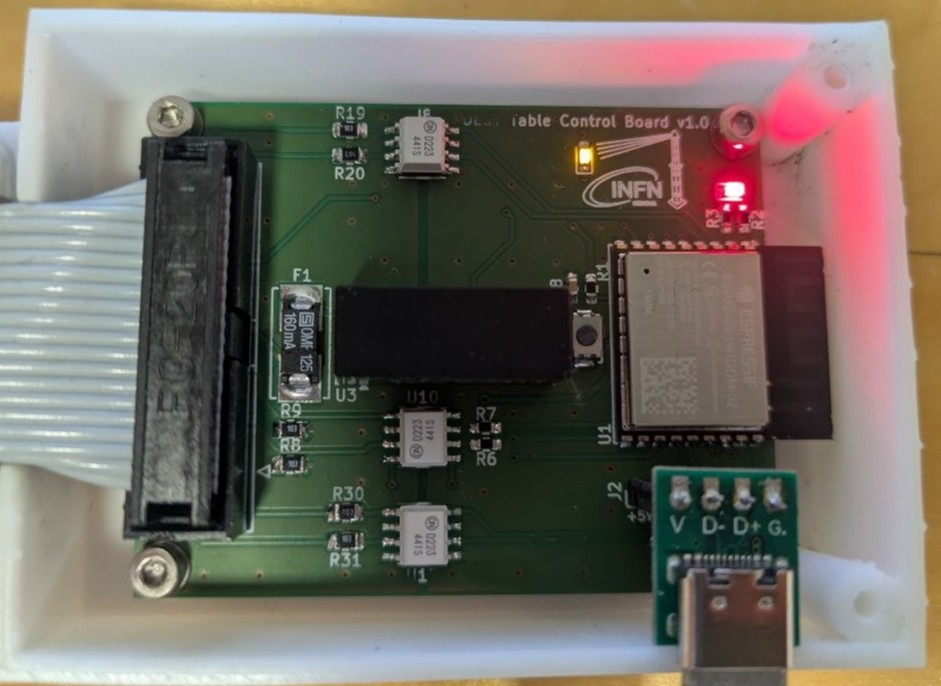}
        \caption{Photo of the assembled device.}
        \label{fig:desy_ctrl_board_3d}
    \end{subfigure}
    \caption{PCB representation of the designed device: (a) traced PCB, (b) photo of the assembled device.}
    \label{fig:desy_ctrl_pcb_combined}
\end{figure}

The idea of the design is to make the device attachable directly to the DESY control panel connector without any external cables, requiring a compact a lightweight solution to avoid any mechanical stress on the connector. The final design foresees a 57.5~mm~$\times$~45~mm 2-layers PCB with some extra off-board space for the SoM's PCB-antenna, which extends beyond the main PCB area. The PCB layout is shown on Fig.~\ref{fig:desy_ctrl_pcb_combined}, left panel. The component placement is done on the both sides of the PCB to make it as compact as possible. Two separate ground planes are present, the first being the control panel's ground (0~V) connected to the 28BSM connector and used for the opto-isolator circuits from the connector side, and the other being the device's own ground (GND) used for the power supply and the ESP32-C3-WROOM-2 SoM. The ground planes are decoupled by a high-voltage capacitor to ensure that the device's ground is not affected by the noise from the control panel's ground. A picture of the assembled PCB is shown in Fig.~\ref{fig:desy_ctrl_pcb_combined}, right panel. To interact with the control panel the PCB has an IDC-20 connector (J1) which is connected to the 28BSM mating connector via a short ribbon cable. The connector pinout is shown in Tab.~\ref{tab:idc_pinout}. 

The final device price is within 100\euro \, including PCB manufacturing and case 3D printing. This low cost allows for a very convenient adoption of the $\texttt{NA64-DTC}$ device in any PS/SPS experimental setup. 


\subsubsection{Firmware development}

The controller firmware implements a custom multi-thread application, running under the FreeRTOS kernel~\cite{freertos}. To allow for future expansions, GPIO pin assignments are passed as compile-time flags, keeping the hardware mapping fully decoupled from application logic -- retargeting to a different board revision therefore requires only to change the configuration file without modifying the application source. To facilitate debugging, serial monitoring is configured at 115\,200~baud with the ESP32 exception decoder filter enabled, allowing decoded stack backtraces to be read directly from the monitor console. 

On startup, the firmware initialises the non-volatile storage subsystem, configures all GPIO lines, attaches quadrature encoder interrupt service routines, and launches a periodic LED timer before connecting to the CERN Wi-Fi network. Once an IP address is obtained, a \texttt{HTTP} server is started on port 80 and the device is ready to accept commands.
The REST API exposes the following endpoints: 
\begin{itemize}
    \item \texttt{GET /api/status} returns a JSON object containing encoder counts, motor direction, encoder fault flags for each axis as well as HW and SW E-Stop states; 
    \item \texttt{POST /api/command/hstart} and \texttt{POST /api/command/vstart} accept a signed encoder-count increment in the request body and initiate motion; 
    \item \texttt{POST /api/command/hstop} and \texttt{/vstop} halt the corresponding axis; 
    \item \texttt{GET /api/command/hreset\_revs} and \texttt{/vreset\_revs} zero the position counters; 
    \item \texttt{GET /api/command/estop\_sw} and \texttt{/clear\_estop} activate and deactivate the software E-Stop.
\end{itemize}

\begin{figure}[t]
    \centering
    \includegraphics[width=0.8\linewidth]{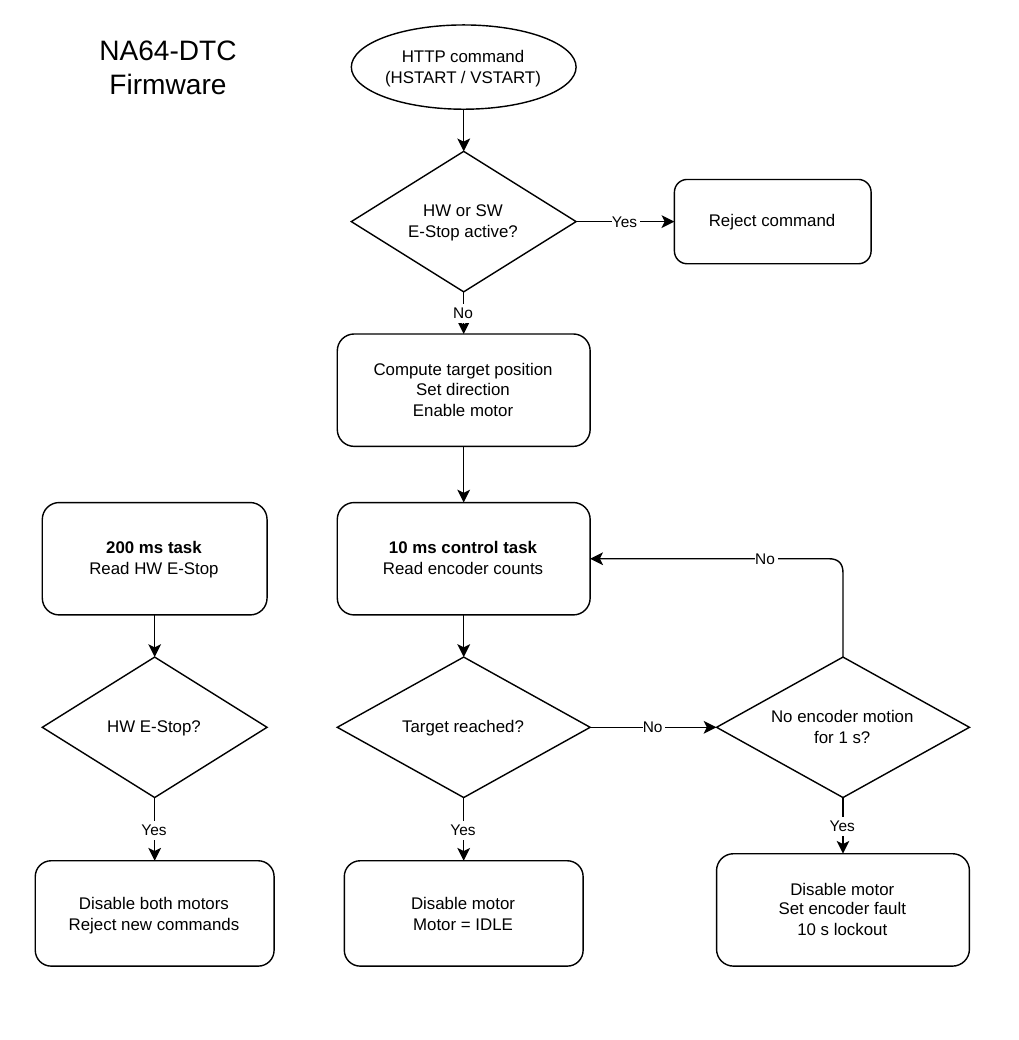}
    \caption{Block diagram of the control algorithm implemented in the \texttt{NA64-DTC} firmware, showing the commands and controls executed once a vertical/horizontal movement is initiated via the \texttt{HTTP REST API}.}
    \label{fig:NA64DTC_firmware}
\end{figure}

Two independent motor axes are implemented: horizontal (left/right) and vertical (up/down), each driven by a direction-plus-enable GPIO triplet. Closed-loop position control relies on quadrature incremental encoders. All four encoder channels share a single Interrupt Service Routine (ISR) that reads the full GPIO input register atomically and applies a 16-entry lookup table for $\times$4 quadrature decoding. A dedicated FreeRTOS task running at a 10~ms period propagates ISR pulse counts into signed absolute position counters, checks whether the target count has been reached, and monitors for motor stalls. Normally, after receiving \texttt{/hstart} or \texttt{/vstart} command the task starts the selected motor and stops it when the monitored counts value reaches the target one or if it gets the corresponding \texttt{stop} command before.  
If a motor is commanded but no encoder movement in the expected direction is detected for one second, a fault is declared, the drive is cut, and a 10-second lockout is applied before the error is automatically cleared. A simplified block diagram of the control algorithm implemented in the firmware is reported in Fig.~\ref{fig:NA64DTC_firmware}.

Safety is enforced at two levels. A hardware E-Stop input from the Emergency Stop button of the DESY Table control panel, polled every 200~ms, immediately disables both drives and prevents any new motion command; its state is reported via the status API. Also, a software E-Stop can be asserted and cleared through the API, with the constraint that clearing is blocked while the hardware E-Stop is asserted. 

A single status LED encodes the system state through distinct blink patterns: fast toggling (100~ms period) for any E-Stop condition, 200~ms period for an encoder fault, asymmetric duty cycles for forward and reverse motion, a 100~ms pulse heartbeat every 1 seconds during network initialisation, and a 100~ms pulse heartbeat every 5 seconds in the idle state.

\subsubsection{User interface software development}

The generic REST API exposed by the built-in \texttt{HTTP} server allows to easily integrate the \texttt{NA64-DTC} in any experiment control system. To allow the immediate use of the device in any setup, two specific users interfaces have been developed, a lightweight browser-based dashboard served directly by the microcontroller for local access, and an EPICS Input/Output Controller (softIOC) running on a different server.

\begin{figure}[t]
  \centering
  \includegraphics[width=0.75\textwidth]{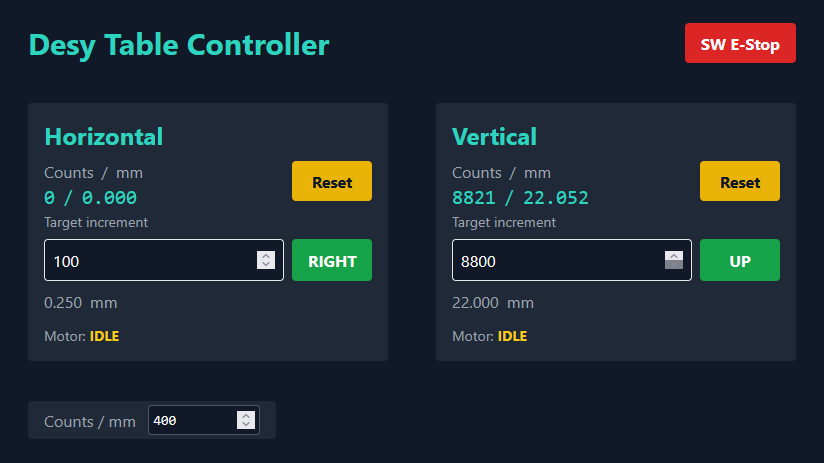}
  \caption{Browser-based web interface of the \texttt{NA64-DTC}, served directly by the embedded \texttt{HTTP} server.}
  \label{fig:webui}
\end{figure}

The web interface is a self-contained single-page HTML/Javascript application, served directly by the ESP32 \texttt{HTTP} server\footnote{To avoid storing the HTML as a separate file in flash, a Python utility is run automatically at build time to convert the web-page html file into a C character array inside the application sources, which are then compiled into the firmware image.}. The interface, shown in Fig.~\ref{fig:webui}, presents two symmetrical panels, one per axis, each showing the current position in encoder counts, a numeric increment input, a directional motion button whose label (RIGHT/LEFT or UP/DOWN) updates dynamically to reflect the sign of the entered value, a reset button, and a stall-error indicator. The page polls \texttt{/api/status} every 400~ms and updates all displayed values without reloading. When a motor is running, its directional button converts to a STOP button; on completion the original label is restored. A software E-Stop button is always visible in the header; activating it reveals a clear button that requires a deliberate two-second press-and-hold to prevent accidental release. If the hardware E-Stop is asserted, all interactive controls are disabled, the clear button is explicitly locked, and a pulsing red banner identifying the active E-Stop type is displayed at the top of the page. Encoder fault conditions are signalled by amber blinking of the affected position readout.

In parallel, a custom SoftIOC was developed to allow the \texttt{NA64-DTC} device integration within an EPICS-based control system - the first use case was the NA64 control system, based on this architecture. The SoftIOC runs on a dedicated server, and communicates with the microcontroller through the Stream Device module over a raw TCP socket established with \texttt{drvAsynIPPortConfigure}. The StreamDevice protocol file defines all \texttt{HTTP}/1.1 transactions: the status query issues a \texttt{GET /api/status} request and parses the JSON response body in a single \texttt{in} format string, extracting both E-Stop flags, encoder counts, motor directions, and encoder error flags for both axes. 
Motor start commands encode the signed encoder-count increment as a zero-padded 32-bit hexadecimal integer in the JSON request body and send the corresponding \texttt{POST} request. Separate protocol entries handle horizontal and vertical stop, encoder reset, and software E-Stop assertion and clearing. 

An EPICS database file defines the full list of Process Variables (PVs) - a full description is given in Table~\ref{tab:pvs} in the Appendix. A single \texttt{longin} master poll record scans at 1~Hz and drives all status updates passively. 
Per-axis position is exposed in physical units through a \texttt{calc} record that divides the raw encoder count by a configurable scale factor. Soft travel limits are implemented  for both axes and enforced in software by \texttt{calcout} validation records that gate motion commands: an absolute-position \texttt{Goto} command is only forwarded to the motor-start record if the target lies within the configured limit range; 
a relative \texttt{Increment} command likewise checks that the resulting position would remain in range. 
If neither check passes, a \texttt{bi} range-alarm record is set to MAJOR severity. 
The unit-to-count conversion for both absolute and relative moves is performed by dedicated \texttt{calcout} records before the integer count increment is written to the StreamDevice output record. Two one-touch preset records, \texttt{BeamCenter} and \texttt{BeamOut}, trigger a \texttt{fanout} that simultaneously writes the stored centre or out coordinates to the horizontal and vertical \texttt{Goto} records, initiating a coordinated two-axis move to predefined survey positions. Hardware and software E-Stop states are mirrored as \texttt{bi} records, and a combined \texttt{calc} record computes their logical OR for use by interlock logic elsewhere in the control system. If wireless communication between the EPICS IOC and the ESP32 SOM inside the NA64-DTC is lost during a movement, the platform motion continues until the target is reached, and simultaneously an EPICS alarm is triggered, to make the operator aware of the issue. The alarm event is logged within the slow-controls database. No further operations are possible until the communication error is fixed. Once this occurs, the alarm condition, after the operator acknowledgement, is automatically cleared and EPICS PVs holding the current platform position are updated reading the last values from ESP32.

\begin{figure}[t]
  \centering
  \includegraphics[width=0.6\textwidth]{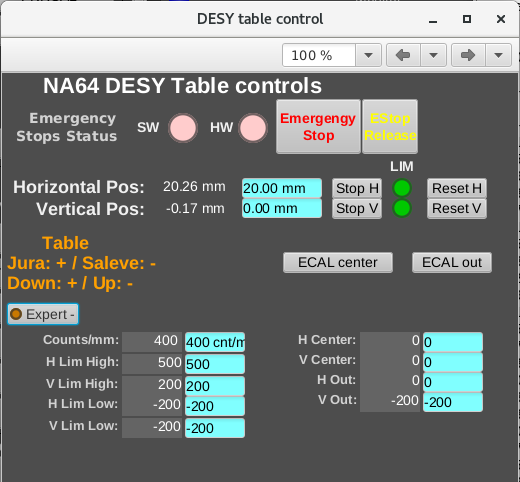}
  \caption{CS-Studio operator panel for the \texttt{NA64-DTC} SoftIOC, in expert view.}
  \label{fig:csstudio_expert}
\end{figure}

To facilitate operator's interaction with the device, a CS-Studio panel was created (see Fig.~\ref{fig:csstudio_expert}). Two display modes are foreseen. In the standard view, the panel solely exposes the controls required for routine operation: software and hardware Emergency Stop status LEDs, a Software \texttt{Emergency Stop} and \texttt{EStop Release} buttons, current horizontal and vertical position readback in millimeters, absolute-position target entry fields for each axis, per-axis \texttt{Stop} and coordinates \texttt{Reset} buttons, range-alarm LEDs. The expert mode also includes the encoder-to-millimeter scale factor (\texttt{mm2cnt}), the stored beam-centre and beam-out coordinates for each axis, and the upper and lower soft travel limits for both axes. Each parameter is presented as a read-back display alongside an editable entry field, allowing authorised operators to adjust the calibration and limit values in situ without restarting the IOC. 

\section{Results}
\subsection{Device test and commissioning}\label{sec:commissioning}

\begin{figure}[t]
  \centering
  \includegraphics[width=0.7\textwidth]{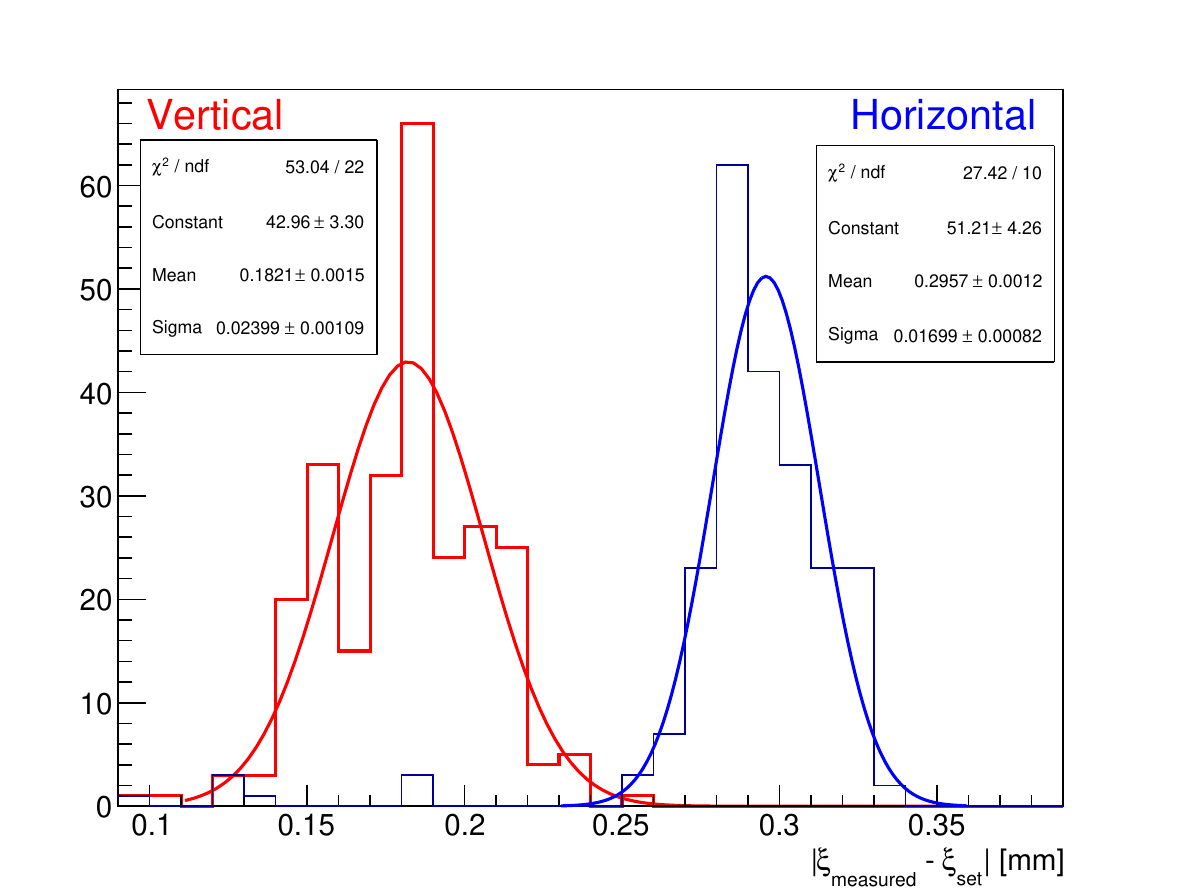}
  \caption{Distribution of absolute difference between DESY table set point $\xi_{set}$ and actual position $\xi_{measured}$ for various movements of the horizontal (blue) and vertical (red) axis, as measured during the NA64 2026 run at H4. The two distributions were fit with a Gaussian probability density function to extract the average accuracy value.}
  \label{fig:fit}
\end{figure}

The \texttt{NA64-DTC} device was commissioned during the two NA64 experimental runs at the PS T9 beamline (February~2026) and at the SPS H4 beamline (April-May~2026). During the PS test, a prototype version of the system was used. After connection to the manual controller of the DESY platform and following a first successful electrical check of the device, a connectivity test was performed, demonstrating proper communication through the SoM Wi-Fi interface. Next, the conversion factor between the platform linear displacement and the encoder counts was validated by comparing the encoder readout with the position displayed by the original DESY Table controller for a series of known platform movements. From the nominal system configuration, the expected conversion coefficient is 400 counts/mm, obtained from the combination of 100 encoder pulses per motor revolution, a mechanical transmission corresponding to 1 mm of table displacement per motor revolution, and the $\times 4$ quadrature decoding implemented in the firmware. The experimentally measured value was 400 counts/mm for both axes, thus fully confirming the expected conversion factor.
The integration test also allowed us to estimate the accuracy of the system, by setting the table position to a defined absolute value and then comparing this with the reported position from the encoder data. Values in range $0.2-0.3$~mm were observed for both axes, across the whole movement range. This estimate was quantitatively confirmed by the analysis of the various DESY platform repositioning performed during the NA64 run at H4, discussed in the next Section. The positioning accuracy was determined by analysing the EPICS logs of the platform movements, comparing for each the target position $\xi_{set}$ with the actual reached position $\xi_{measured}$, and considering the distribution of the absolute difference, independently for each axis. The resulting distribution, reported in Fig.~\ref{fig:fit}, was then fitted with a Gaussian probability density function to extract the mean value and its uncertainty. For the horizontal (vertical) axis, a value of $(295\pm1)~\mu$m ($(182\pm1)~\mu$m) was found.

The observed positioning accuracy is entirely attributable to the finite inertia of the DESY table. When the target position is reached, the corresponding motor is disabled by the \texttt{NA64-DTC} with an intrinsic control accuracy of approximately one encoder count ($0.01$~mm). Owing to the large mass of the moving platform, however, the motion does not stop instantaneously after motor deactivation, resulting in a residual displacement consistent with the observed positioning accuracy. This value represents the ultimate alignment performances of the DESY platform, being a factor $\approx 2$ better then the accuracy reached when operating the device with the manual controller. We observe that this accuracy is fully adequate for the detector-alignment and calibration operations for which the DESY Table is typically employed. In the specific NA64 application, the ECAL cells have transverse dimensions of $3.8 \times 3.8~\mathrm{cm}^2$, while the H4 electron beam has a typical spot size of about 1.5 cm (FWHM), so the measured positioning uncertainty corresponds to less than 1$\%$ of the cell width. Similar centimetre-scale transverse dimensions are common in calorimeter-based beam-test and calibration setups, making a sub-millimetre positioning accuracy sufficient for reproducibly placing the detector with respect to the beam. This mechanical positioning accuracy should not be confused with the precision with which the beam particle impact point may be reconstructed during offline analysis. When a suitable tracking system is installed upstream of the detector, the impact position can be determined event by event with substantially better precision, independently of the absolute positioning accuracy of the motorized platform.

\begin{figure}[t]
  \centering
  \includegraphics[width=0.6\textwidth]{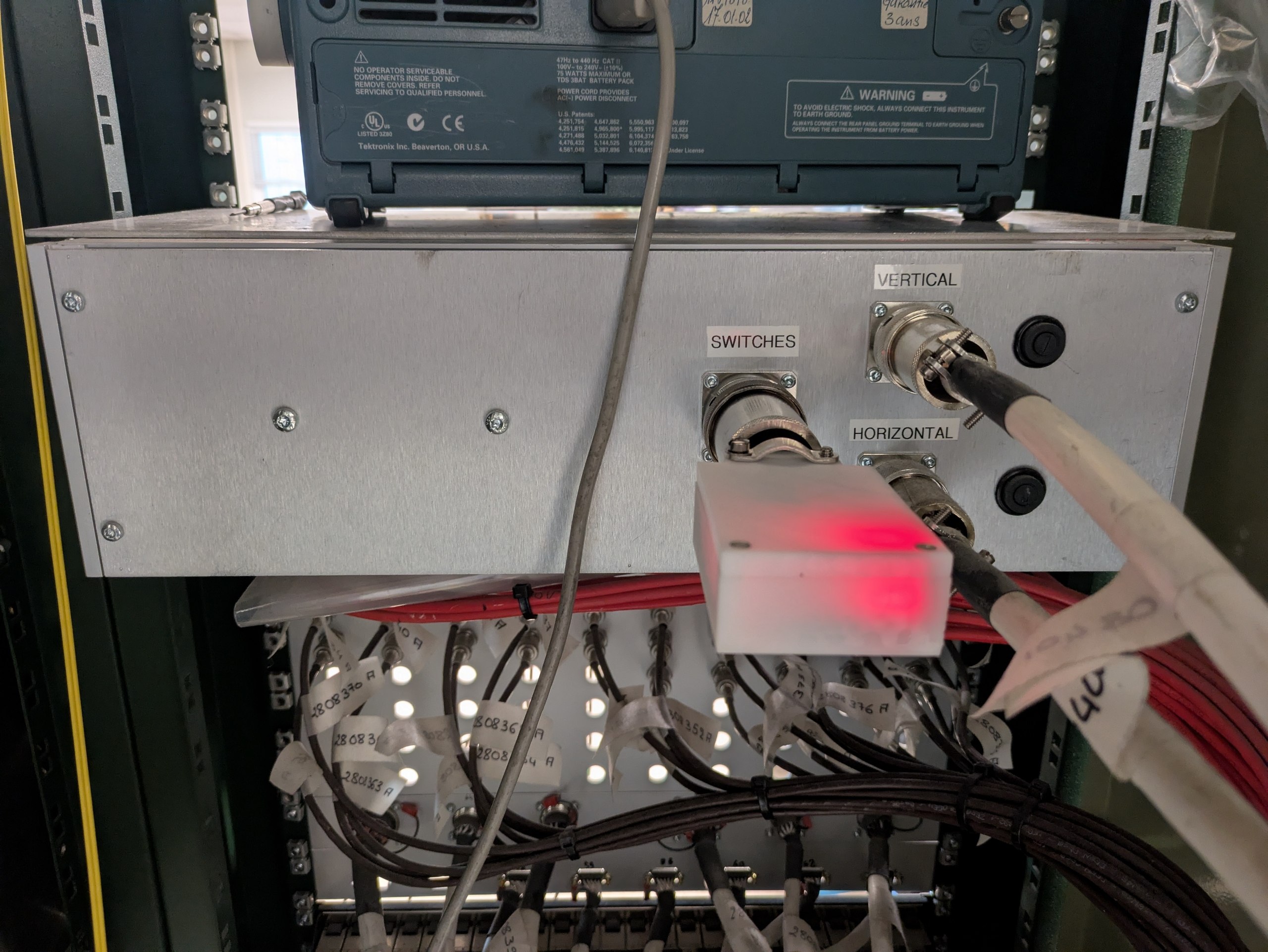}
  \caption{Rear view of the PS T9 DESY Table control panel with the \texttt{NA64-DTC} device installed.}
  \label{fig:T9_ctrl_panel_back}
\end{figure}

A follow-up test at the H4 beamline was performed using the final device together with a different manual controller for the DESY platform. The test again confirmed the proper functionality of the system, and similar performances in terms of movement accuracy were measured. An undocumented feature of the controller, namely the possibility that the movement axes are inverted with respect to the nominal configuration, was observed. A corresponding modification of the device firmware was therefore introduced to account for this behaviour, ensuring compatibility with both controller versions. Since, as discussed in the next Section, the device was continuously used for a period of about two months, the H4 test also allowed to assess the reliability of the system for an extended working time. During the whole run, no operational issues were observed, and no manual interventions to the device were required. 

\subsection{Operational experience in NA64}\label{sec:NA64}

The successful commissioning of the device motivated its use during the whole NA64 2026 run at H4 - the \texttt{NA64-DTC} was routinely used in the experiment to move the ECAL detector position with respect to the beamline for calibration purposes and for proper alignment with the nominal beam direction. A quantitative summary of the device operations is reported in Tab.~\ref{tab:runConditions}. 

\begin{table}[h!]
\centering
\caption{Summary of \texttt{NA64-DTC} operation during the NA64 2026 run at the CERN H4 beamline.}
\label{tab:runConditions}
\begin{tabular}{lcc}
\hline
\textbf{Run parameter} & \textbf{Horizontal} & \textbf{Vertical} \\
\hline
Motion commands executed & 307 & 329 \\
Total travelled distance  & 8.987 m & 13.888 m \\
\hline
Wi-Fi disconnection events & \multicolumn{2}{c}{None} \\
Encoder fault events       & \multicolumn{2}{c}{None} \\
Continuous operation time  & \multicolumn{2}{c}{50 d 09 h 50 min 19 s} \\
\hline
\end{tabular}
\end{table}

Specifically, in NA64 the DESY platform is mostly used for ECAL calibration task, involving many changes of the calorimeter position along the beam-line so that the beam sequentially impinges on the centre of each of the cells.  For each beam position, the cell response is measured and compared with Monte Carlo simulations to determine the corresponding energy calibration constant~\cite{NA64:2025rib}. When the calibration task is performed operating the DESY table in manual mode, the platform is routinely repositioned by the operator through the control panel, acting on the corresponding push buttons until the desired detector position is met, as defined by reading the visual position indicators. Besides increasing the reconfiguration time, this procedure is also a potential source of operational mistakes, such as incorrect detector positioning or deviations from the intended calibration pattern. Furthermore, in this mode neither records nor logs of the platform movements are kept, thus relying on manual book-keeping to keep proper track of the detector positioning. 
The use of the \texttt{NA64-DTC} enables to circumvent these limitations, allowing the entire calibration sequence to be automated via the experiment slow-controls system. Detector positions can be pre-programmed and executed autonomously via EPICS, ensuring reproducible operation, in a shorter time. 
The benefits of this approach were observed during NA64 data-taking periods. Concerning repositioning errors, in the last (2025) ECAL calibration campaigns performed with the DESY Table operated manually, positioning mistakes occasionally occurred, requiring careful online verification of the data to eventually repeat a mispositioned cell characterization.
For one scan, a cell was inadvertently excluded from the scan, without the operator being able to acknowledge this, requiring an ad-hoc offline analysis procedure. 
During the 2026 run, in which a total of seven detector scans were performed  using the \texttt{NA64-DTC}, no positioning errors of this kind were observed. 
Concerning the gain in repositioning time, a typical ECAL calibration scan requires an accumulated statistics of about 20 SPS spills for each of the 30 ECAL channels, resulting in a total procedure time of about 3 hours, in the optimal scenario of 4 SPS spills delivered per minute. When the DESY platform is operated in manual mode, each detector repositioning requires approximately from 30 s to 1 minute per cell, largely due to the operator intervention\footnote{The reported value represents our estimate based on several years of NA64 operations.}. For a complete ECAL calibration, this corresponds to a total time of up to 30 minutes, i.e. to a non-negligible fraction of the procedure overall duration. The use of the \texttt{NA64-DTC} devices allows to reduce such time to a significantly smaller value, since in this mode the repositioning solely corresponds to the actual time required for the platform movement. For the NA64 ECAL detector, whose cells have transverse dimensions of a few centimetres, the measured time for a typical cell-to-cell repositioning is $(15.8 \pm 0.9)$~s, where the uncertainty is largely determined by the 1 Hz scan frequency of the EPICS PVs recording the platform position. Comparable results are expected for other detector R\&D activities involving calorimeters calibration.

Another gain in operational efficiency is provided by the \texttt{NA64-DTC} during detectors installation and early commissioning. The DESY Table controller does not provide local movement controls on the platform itself, thus requiring a first operator to remain at the control panel -- typically located inside the experiment control room -- while a second operator verifies the detector position inside the experimental area. As a result, initial alignment procedures require continuous coordination between two people. By enabling remote operation through a laptop computer, the \texttt{NA64-DTC} allows a single operator to directly observe the detector position while simultaneously controlling the platform motion. This capability simplifies the detector installation and alignment procedures, reduces manpower requirements, and shortens commissioning activities.

\section{Discussion and Conclusions}\label{sec:conclusions}
We have described the design, development, and commissioning of a compact automation controller \texttt{NA64-DTC} for the so-called ``DESY Table'' motorised positioning platform routinely used for CERN North Area and East Area fixed target experiments. The device has been originally developed for the NA64 experiment operating at the H4 beamline, but, thanks to use of a generic \texttt{HTTP}-based interface and the absence of any mechanical or electrical requirement, it can be potentially used by any other experiment.

The system was designed to interface non-invasively with the existing platform hardware, requiring neither modifications to the original control electronics nor an external power supply. In addition to basic motion control, the system supports configurable positioning parameters, software travel limits, predefined motion presets, and dedicated emergency stop functionality. A lightweight web interface is provided to allow for immediate use of the device, while experiment-specific integration solutions can be easily developed by interfacing with the REST API exposed by the \texttt{HTTP} server.

The controller was successfully commissioned during operation at the PS T9 and SPS H4 beamlines in 2026, where it was used during ECAL calibration activities. The performed tests demonstrated stable operation of the device under realistic experimental conditions, including reliable Wi-Fi communication, correct encoder-based position tracking, and robust behaviour of the safety logic throughout extended periods of continuous use.

The operational experience gained during the 2026 NA64 run demonstrated the reliability and usefulness of the \texttt{NA64-DTC} in routine experimental operation. The device enabled automated detector positioning through integration with the experiment control system, reducing manual intervention during calibration and alignment procedures while improving their reproducibility. Owing to the widespread use of the DESY Table platform at CERN beamline facilities, the proposed solution may be of interest to other fixed-target experiments and detector R$\&$D activities.

\section*{Author Contributions}
All authors have read and agreed to the published version of the manuscript.

\section*{Funding}
This work is part of a project that has received funding from the European Research Council (ERC) under the European Union's Horizon 2020 research and innovation programme, Grant Agreement No. 947715 (POKER). This project has received funding from the European Union's Horizon Europe Research and Innovation programme under Grant Agreement No. 101057511 (EURO-LABS).

\section*{Data Availability}
All the device electrical schematics, PCB design, and firmware sources are available from a Zenodo-hosted dataset~\cite{celentano2025_na64dtc}.

\section*{Acknowledgments}
We gratefully acknowledge the support of the CERN management and staff, in particular S. Girod, J. Lendaro and V. Marchand, for their help with the original DESY platform documentation, as well as the personnel of the INFN-Genova electronic workshop, in particular A. Rovani, for the support given during the design and construction of the \texttt{NA64-DTC} device. This result is part of a project that has received funding from the European Research Council (ERC) under the European Union's Horizon 2020 research and innovation programme, Grant Agreement No. 947715 (POKER).


\clearpage
\appendix
\section{Technical details for the motorized platform controller}\label{app:techDetails}

\begin{table}[h]
\caption{Pinout of the 28BSM replication connector of the DESY table manual controller}
\label{tab:28bsm_pinout}
\centering
\resizebox{\textwidth}{!}{
\begin{tabular}{ll}
\toprule
\textbf{Pin(s)} & \textbf{Description} \\
\midrule
1, 4, 15, 18 & 0 V ground connections \\
2 & Horizontal axis +24VDC power supply from DESY Table (horizontal 19BSM connector) \\
3 & ``Right'' button signal \\
6 & ``Left'' button signal \\
7 & Horizontal axis ``Fast'' button signal \\
8 & Horizontal axis ``Slow'' button signal \\
9 & Horizontal axis encoder B output \\
10 & Horizontal axis encoder A output \\
11 & Horizontal axis drive enable signal \\
12 & Emergency stop signal \\
13 & +24VDC power supply from horizontal 19BSM connector \\
16 & Vertical axis +24VDC power supply from DESY Table (vertical 19BSM connector) \\
17 & ``Down'' button signal \\
20 & ``Up'' button signal \\
21 & Vertical axis ``Fast'' button signal \\
22 & Vertical axis ``Slow'' button signal \\
23 & Vertical axis encoder A output \\
24 & Vertical axis encoder B output \\
25 & Vertical axis drive enable signal \\
5, 14, 19, 26--28 & Not used \\
\bottomrule
\end{tabular}
}
\end{table}   

\begin{center}
\begin{longtable}{p{0.28\textwidth}p{0.12\textwidth}p{0.06\textwidth}p{0.44\textwidth}}
\caption{Process Variables of the \texttt{NA64desytable} EPICS database. The full PV name is obtained by prepending the macro \texttt{\$(P)} (instantiated as \texttt{NA64:desytable}) to each suffix listed below.}
\label{tab:pvs} \\
\toprule
\textbf{PV suffix} & \textbf{Record type} & \textbf{R/W} & \textbf{Description} \\
\midrule
\endfirsthead
\toprule
\textbf{PV suffix} & \textbf{Record type} & \textbf{R/W} & \textbf{Description} \\
\midrule
\endhead
\bottomrule
\endfoot

\multicolumn{4}{l}{\textit{Polling}} \\[2pt]
\texttt{:StatusPoll}
    & \texttt{longin}  & R  & Master poll record; queries \texttt{GET /api/status} at 1~Hz and fans out to all status PVs \\[6pt]

\multicolumn{4}{l}{\textit{E-Stop status}} \\[2pt]
\texttt{:HwEStop}
    & \texttt{bi}      & R  & Hardware E-Stop state (0\,=\,Released, 1\,=\,ACTIVE) \\
\texttt{:SwEStop}
    & \texttt{bi}      & R  & Software E-Stop state (0\,=\,Released, 1\,=\,ACTIVE) \\
\texttt{:EStop}
    & \texttt{calc}    & R  & Global E-Stop flag: logical OR of \texttt{HwEStop} and \texttt{SwEStop} \\[6pt]

\multicolumn{4}{l}{\textit{E-Stop commands}} \\[2pt]
\texttt{:SwEStopOn}
    & \texttt{bo}      & W  & Activates software E-Stop (\texttt{GET /api/command/estop\_sw}) \\
\texttt{:SwEStopClear}
    & \texttt{bo}      & W  & Clears software E-Stop (\texttt{GET /api/command/clear\_estop}) \\[6pt]

\multicolumn{4}{l}{\textit{Axis readbacks}} \\[2pt]
\texttt{:H:Counts / :V:Counts}
    & \texttt{ai}      & R  & Raw encoder count for horizontal / vertical axis (encoder pulses) \\
\texttt{:H:Direction / :V:Direction}
    & \texttt{mbbi}    & R  & Motor direction: H axis (0\,=\,IDLE, 1\,=\,RIGHT, 2\,=\,LEFT); V axis (0\,=\,IDLE, 1\,=\,UP, 2\,=\,DOWN) \\
\texttt{:H:EncError / :V:EncError}
    & \texttt{bi}      & R  & Encoder fault flag for horizontal / vertical axis (0\,=\,OK, 1\,=\,ERROR) \\
\texttt{:H:Pos / :V:Pos}
    & \texttt{calc}    & R  & Current position in mm for horizontal / vertical axis (\texttt{Counts\,/\,mm2cnt}) \\[6pt]

\multicolumn{4}{l}{\textit{Motion commands}} \\[2pt]
\texttt{:H:Goto / :V:Goto}
    & \texttt{ao}      & W  & Absolute move target in mm for horizontal / vertical axis; triggers limit check on write \\
\texttt{:H:Increment / :V:Increment}
    & \texttt{ao}      & W  & Relative move increment in mm for horizontal / vertical axis; triggers limit check on write \\
\texttt{:H:Start / :V:Start}
    & \texttt{longout} / \texttt{ao} & W  & Sends signed encoder-count increment to \texttt{POST /api/command/hstart} or \texttt{vstart} \\
\texttt{:H:Stop / :V:Stop}
    & \texttt{bo}      & W  & Stops horizontal / vertical motor (\texttt{POST /api/command/hstop} or \texttt{vstop}) \\
\texttt{:H:Reset / :V:Reset}
    & \texttt{bo}      & W  & Resets horizontal / vertical encoder counter (\texttt{GET /api/command/hreset\_revs} or \texttt{vreset\_revs}) \\[6pt]

\multicolumn{4}{l}{\textit{Internal logic}} \\[2pt]
\texttt{:H:CheckLim / :V:CheckLim}
    & \texttt{calcout} & -- & Validates absolute target against soft limits; forwards to \texttt{TargetCalc} if in range \\
\texttt{:H:CheckIncLim / :V:CheckIncLim}
    & \texttt{calcout} & -- & Validates current position plus increment against soft limits; forwards to \texttt{IncCalc} if in range \\
\texttt{:H:UpdateRangeAlarm / :V:UpdateRangeAlarm}
    & \texttt{calcout} & -- & Sets \texttt{RangeAlarm} if both limit checks fail \\
\texttt{:H:RangeAlarm / :V:RangeAlarm}
    & \texttt{bi}      & R  & Range alarm for horizontal / vertical axis (0\,=\,Ready, 1\,=\,OUT OF RANGE, MAJOR severity) \\
\texttt{:H:TargetCalc / :V:TargetCalc}
    & \texttt{calcout} & -- & Converts absolute target in mm to encoder-count increment; writes to \texttt{Start} \\
\texttt{:H:IncCalc / :V:IncCalc}
    & \texttt{calcout} & -- & Converts relative increment in mm to encoder counts; writes to \texttt{Start} \\[6pt]

\multicolumn{4}{l}{\textit{Preset moves}} \\[2pt]
\texttt{:BeamCenter / :BeamOut}
    & \texttt{bo}      & W  & Triggers coordinated two-axis move to beam-centre / beam-out position \\
\texttt{:BeamCenter:Fan / :BeamOut:Fan}
    & \texttt{fanout}  & -- & Forwards preset trigger to horizontal and vertical \texttt{Goto} records \\
\texttt{:BeamCenter:H / :BeamOut:H}
    & \texttt{calcout} & -- & Copies \texttt{H:Center} / \texttt{H:Out} setpoint to \texttt{H:Goto} \\
\texttt{:BeamCenter:V / :BeamOut:V}
    & \texttt{calcout} & -- & Copies \texttt{V:Center} / \texttt{V:Out} setpoint to \texttt{V:Goto} \\[6pt]

\multicolumn{4}{l}{\textit{Configuration and soft limits}} \\[2pt]
\texttt{:mm2cnt}
    & \texttt{ai}      & R/W & Encoder scale factor (counts/mm) \\
\texttt{:H:LimL / :V:LimL}
    & \texttt{ai}      & R/W & Lower soft travel limit in mm for horizontal / vertical axis \\
\texttt{:H:LimH / :V:LimH}
    & \texttt{ai}      & R/W & Upper soft travel limit in mm for horizontal / vertical axis \\
\texttt{:H:Center / :V:Center}
    & \texttt{ai}      & R/W & Stored beam-centre coordinate in mm for horizontal / vertical axis \\
\texttt{:H:Out / :V:Out}
    & \texttt{ai}      & R/W & Stored beam-out coordinate in mm for horizontal / vertical axis \\

\end{longtable}
\end{center}

\bibliographystyle{unsrt}
\bibliography{biblio}

\end{document}